\documentclass[prd,
preprintnumbers,amsmath,amssymb]{revtex4}
\usepackage{graphicx}
\usepackage{amsmath}
\DeclareGraphicsExtensions{.pdf,*.jpg}
\DeclareMathOperator\arctanh{arctanh}

 \newcommand{\la}[1]{\label{#1}}

 \def\2{\frac{1}{2}}
 \def\4{\frac{1}{4}}
 
  \def\Rc{\mathcal{R}}
 
 \def \gta {\mathrel{\vcenter
      {\hbox{$>$}\nointerlineskip\hbox{$\sim$}}}}

\newcommand{\beq}{\begin{equation}}
\newcommand{\eeq}{\end{equation}}

\newcommand{\bea}{\begin{eqnarray}}
\newcommand{\eea}{\end{eqnarray}}
\newcommand{\be}{\begin{equation}}
\newcommand{\ee}{\end{equation}}

\begin{document}

\title{Exact solutions for Vacuum Decay in Unbounded Potentials}

\author{N. Tetradis 
}
\email{ntetrad@phys.uoa.gr}
\affiliation{ 
Department of Physics,
University of Athens,
University Campus,
Zographou 157 84, Greece}

\begin{abstract}
The Standard Model Higgs potential may become unbounded from below 
at large field values 
$h \gta h_{\rm top} \sim 10^{10}\,{\rm GeV}$, with
important cosmological implications.
For a potential of this form, the commonly assumed scenario of a nucleated thin-wall bubble driving
the transition from the electroweak vacuum to the 
unstable region does not apply. 
We present exact analytical solutions for potentials that have the same qualitative form as the 
Higgs potential.
They show that the transition is driven by a thick-wall
spherical bubble of true vacuum, with a surface that expands at 
asymptotically the speed of light. A 
`crunch' singularity appears in the quasi-AdS interior, with the collapsed region also   
expanding at asymptotically the speed of light. The singularity is surrounded by a region of 
trapped surfaces whose boundary forms an apparent horizon. 
An event horizon separates the singularity from the bubble exterior, so that
the expansion of the bubble surface is not affected by the collapse of the 
interior. 
The solutions provide exact descriptions of the geometry for thick-wall bubbles and 
are consistent with the analysis of \cite{Higgstory,Repeat} for the Higgs potential.

\end{abstract}

\maketitle

\section{Introduction}\la{intro}

The Standard Model Higgs potential may display an instability 
at large field values 
$h \gta h_{\rm top} \sim 10^{10}\,{\rm GeV}$~\cite{1112.3022,1307.3536,1507.08833}, with
important cosmological implications~\cite{1210.6987,1301.2846,1503.05193,Higgstory,1605.04974,1606.00849,1608.02555,1608.08848,1706.00792,1809.06923,2011.03763}.
After radiative corrections are taken into account, the SM potential can be approximated 
through an effective running coupling as
\beq\label{eq:VhSM}
V \approx -\lambda(h)\frac{h^4}{4} ,\qquad
\lambda(h)=b \ln \frac{h^2}{e^{1/2}h_{\rm top}^2}\qquad\hbox{with}\qquad  b \approx \frac{0.15}{(4\pi)^2} .
\eeq
The electroweak vacuum is located at $h_{\rm false}\approx 0$. It is separated by a barrier 
located at $h=h_{\rm top}$ from a region where the potential would become unbounded from
below unless gravitational effects generate a new deep minimum around the Planck scale. 
%
For a potential of this form, the commonly assumed scenario of a thin-wall bubble driving
the transition from the false vacuum to the unstable region of the potential does not apply. 
After the nucleation, the field within the bubble does not take a constant value, but instead `rolls' down 
the potential. The situation becomes more complicated when gravity is taken into 
account. Regions of constant negative energy density can be described in terms of the
anti-de Sitter (AdS) spacetime, leading to the Coleman-de Luccia picture of vacuum 
decay \cite{Coleman}.
However, if the field evolves in time, the interior is not captured sufficiently well
by the standard AdS geometry.
Moreover, the presence of a dynamical field can lead to the appearance of a `crunch'
singularity \cite{Coleman,Higgstory}.

The combination of these factors has raised doubts even for the validity of the basic 
conclusion that a sufficiently large bubble will expand forever after its nucleation.  
In \cite{Riotto} it was argued that expanding bubbles nucleated during inflation
eventually turn around and collapse into black holes because of the rapid evolution of the
Higgs field in the interior towards larger values. 
In \cite{Repeat} it was shown that this conclusion relies on the assumption made in \cite{Riotto}
that the interior of the bubble is homogeneous, an approximation that is not valid in the
exact solution. The numerical study of the equations of motion (eom) has shown that a
`crunch' singularity appears in the interior of the bubble, but the bubble surface keeps 
expanding for sufficiently large bubbles \cite{Repeat}. In this sense, the original
analysis of \cite{Higgstory} remains valid and the Hubble scale during inflation must be
constrained in order for the catastrophic bubble nucleation to become improbable.

The purpose of the present paper is to present further evidence 
in order to confirm the above conclusion.
The analysis of \cite{Higgstory} made use of the thin-wall approximation, while 
\cite{Repeat} relied heavily on   
a numerical solution. In this respect, an analytical treatment beyond the thin-wall 
approximation would be more transparent. We are especially interested in vacuum decay
during inflation. 
Earlier work on the various saddle points
associated with vacuum decay in de Sitter (dS) spacetime \cite{EWeinberg} has not
considered the evolution of the spacetime after tunnelling in the case of an unbounded potential. 
Of particular interest is the structure of 
the global geometry, including possible horizons, as well as the nature of the
singularities that may appear. All these would become clearer in a Penrose diagram
constructed through an exact solution.

Unfortunately, it is not easy to find analytical solutions 
for the realistic Higgs potential of eq. (\ref{eq:VhSM}).
However, exact solutions can be derived for potentials that have the same qualitative form.
A particular example is the potential
\be
V(h) = - \frac{1}{4}\lambda h^4
\label{pot1} \ee
with constant $\lambda>0$.
In the absence of gravity, this potential has solutions 
characterized as Fubini instantons \cite{Fubini} of the form
\beq 
h(t_{\rm E},r) = \frac{h_0}{1+(r^2+t_{\rm E}^2)/r_0^2}
\qquad  \hbox{with} \qquad 
r_0= \frac{1}{h_0}\sqrt{\frac{8}{\lambda}}
\label{eq:Fubini}\eeq
and arbitrary $h_0$. The instanton action is 
\be
S_{\rm F}=\frac{8\pi^2}{3\lambda},
\label{actionf} \ee
independently of $h_0$. 
The instantons describe tunnelling from a false vacuum at $h=0$ to a vacuum with an arbitrary 
field value $h_0$. 
The existence of solutions for any value of $h_0$ and the independence of
the action on this parameter are 
reflections of the scale invariance of the theory. 
The logarithmic field dependence of $\lambda$ in the SM potential breaks the scale
invariance and results in the presence of
a barrier. As a result, a particular value of $h_0$  is selected in the unique 
instanton solution. 

When gravity is taken into account, the full Einstein equations must be solved. 
Even though this seems like a formidable task, it is possible to derive analytical 
solutions for particular forms of the potential \cite{Dong,Kanno1,Kanno2,Espinosa1,Espinosa2}.
For potentials that mimic the qualitative form of
eq. (\ref{pot1}), an efficient method is to allow for a conformal coupling of the field
to gravity, find a solution, and eventually transform it to the Einstein frame.
Such an example was presented in \cite{Repeat} in an asymptotically flat spacetime. 
In the following we shall analyze this solution in more detail and also present 
a novel one in an asymptotically de Sitter (dS) spacetime, which is relevant for 
bubble nucleation during inflation. We shall construct Penrose diagrams for the global
spacetimes in order to understand the evolution of the nucleated bubbles, as well as the
nature of the horizons and singularities that appear. 
In all cases we assume that the potential has a locally stable false
vacuum at the origin as a result of small corrections that we neglect. 
Even though our solutions leave the field value $h_0$ at the
center of the bubble undetermined, a particular value is selected when
these modifications are taken into account, similarly to what happens for the SM potential.

We consider a scalar field $h$ with a non-minimal coupling to gravity.
We express all dimensionful quantities in units of the reduced Planck mass
$1/M_{\rm Pl}^2=8 \pi G$. 
The action is
\beq S = \int d^4x \sqrt{|\det g|} 
\left[\frac{1}{2} \Rc - \frac12 g^{\mu\nu}(\partial_\mu h)(\partial_\nu h) 
-\frac{1}{2}\xi \Rc\, h^2- V(h)\right],
\label{actionm}\eeq
where we use the convention $(-,+,+,+)$ for the Lorentzian signature of the metric.
In the context of quantum tunnelling, we also consider the analytic continuation to
Euclidean signature $(+,+,+,+)$, with an inverted potential $-V(h)$. 
The action is 
\beq S = \int d^4x \sqrt{\det g} 
\left[-\frac{1}{2} \Rc + \frac12 g^{\mu\nu}(\partial_\mu h)(\partial_\nu h) 
+\frac{1}{2}\xi \Rc\, h^2+ V(h)\right].
\label{actione} \eeq

\section{Exact solutions in asymptotically flat space}

It is remarkable that the Fubini instanton remains a solution in a fully 
gravitational setting for 
a conformal coupling $\xi=1/6$. The Einstein equations are satisfied with a flat metric, 
while eq. (\ref{eq:Fubini}) solves the eom of the field $h$. 
This observation allows the
derivation of an exact solution for a minimally coupled field in the Einstein
frame with a modified potential. 
A conformal transformation 
\be
g_{\mu\nu}\to \Omega\, g_{\mu\nu}\qquad  \hbox{with} \qquad
\Omega=1-\frac{h^2}{6},
\label{conftrans} \ee
followed by the redefinition of the scalar field as
\be 
\int dh \frac{1}{1-{h^2}/{6}}=\sqrt{6}
\arctanh\frac{h}{\sqrt{6}}
\label{redefinition} 
\ee
so that it has a canonical kinetic 
term, leads to the actions (\ref{actionm}), (\ref{actione}) with $\xi=0$ and
\be
V(h) = -9\lambda  \sinh^4\frac{h}{\sqrt{6}}.
\label{pot2} \ee 
(Notice that we have used the same notation for both the original and the redefined field.)
The solution of the eom for a Lorentzian signature is
\begin{eqnarray} 
h(t,r)&=&\sqrt{6} \arctanh\frac{h_0/\sqrt{6}}{1 + (r^2 - t^2)/r_0^2}
\label{eq:fieldonf} \\
g_{\mu\nu} =A^2(t,r) \eta_{\mu\nu}
&=& \left[1-\frac{h_0^2/6}{(1+(r^2-t^2)/r_0^2)^2}\right] \eta_{\mu\nu},
\label{eq:metriconf}\end{eqnarray}
with arbitrary $h_0$ and
$r_0 =  \sqrt{8/\lambda}/h_0$. 
The continuation to Euclidean signature provides
an $O(4)$-symmetric solution of the eom resulting from the 
action (\ref{actione}) with $\xi=0$ and
potential given by eq. (\ref{pot2}). The on-shell action of this configuration is
again given by eq. (\ref{actionf}). This solution was derived in \cite{Repeat} in order
to support the numerical analysis for the realistic Higgs potential. We summarize here
its main features and display them more clearly through the construction of the
corresponding Penrose diagram.

Eqs. (\ref{eq:fieldonf}), (\ref{eq:metriconf}) 
provide an analytical description of tunnelling in situations
that the potential has a very deep minimum or is totally unbounded from below. 
The Euclidean solution describes tunnelling from a false vacuum at $h=0$ towards the 
lower part of the potential. As we discussed in the previous section, 
in physically interesting situations the scale invariance is broken
by additional terms in the potential that provide a barrier and
lead to the minimization of the action for 
a particular value of $h_0$. The on-shell action, whose leading term is given by
eq. (\ref{actionf}), determines the exponential suppression of the tunnelling 
probability. 

After tunnelling, the system can be described by eqs. (\ref{eq:fieldonf}), 
(\ref{eq:metriconf}), with the initial time taken as $t=0$. 
It is apparent from eq. (\ref{eq:fieldonf}) that the maximal field value at this time,
obtained at $r=0$, becomes infinite for $h_0=\sqrt{6}$. The analysis becomes unreliable 
for $h_0$ close to this maximal value, as the field takes values far 
above the Planck scale, where corrections to the potential are expected. However, 
for $h_0$ sufficiently below $\sqrt{6}$ there is always an 
initial regime in the evolution when the analysis is reliable.

\begin{figure}[t!]
\centering
\vspace{-1cm}
\includegraphics[width=0.35\textwidth]{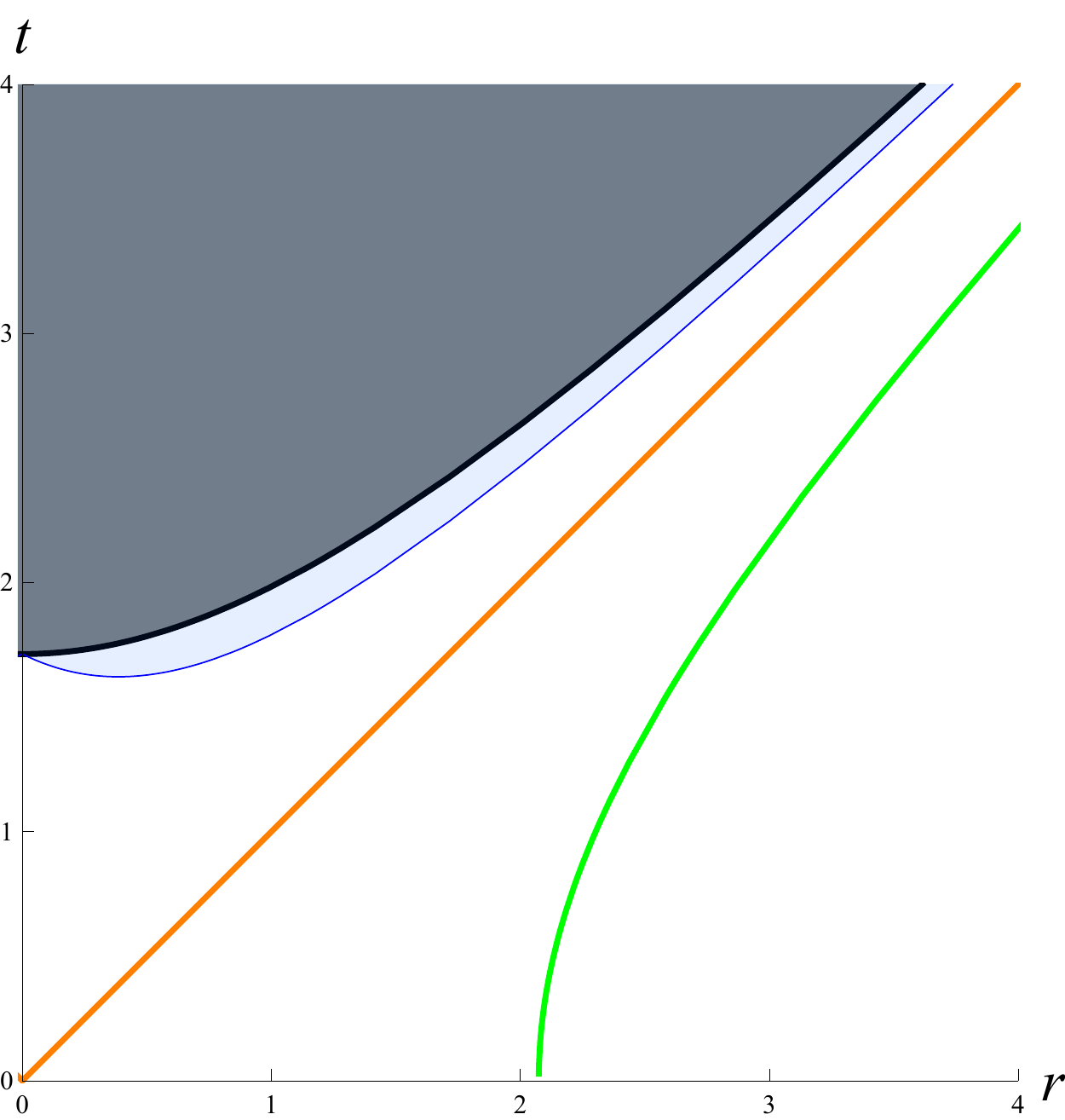} 
\hspace{2cm}
\includegraphics[width=0.4\textwidth]{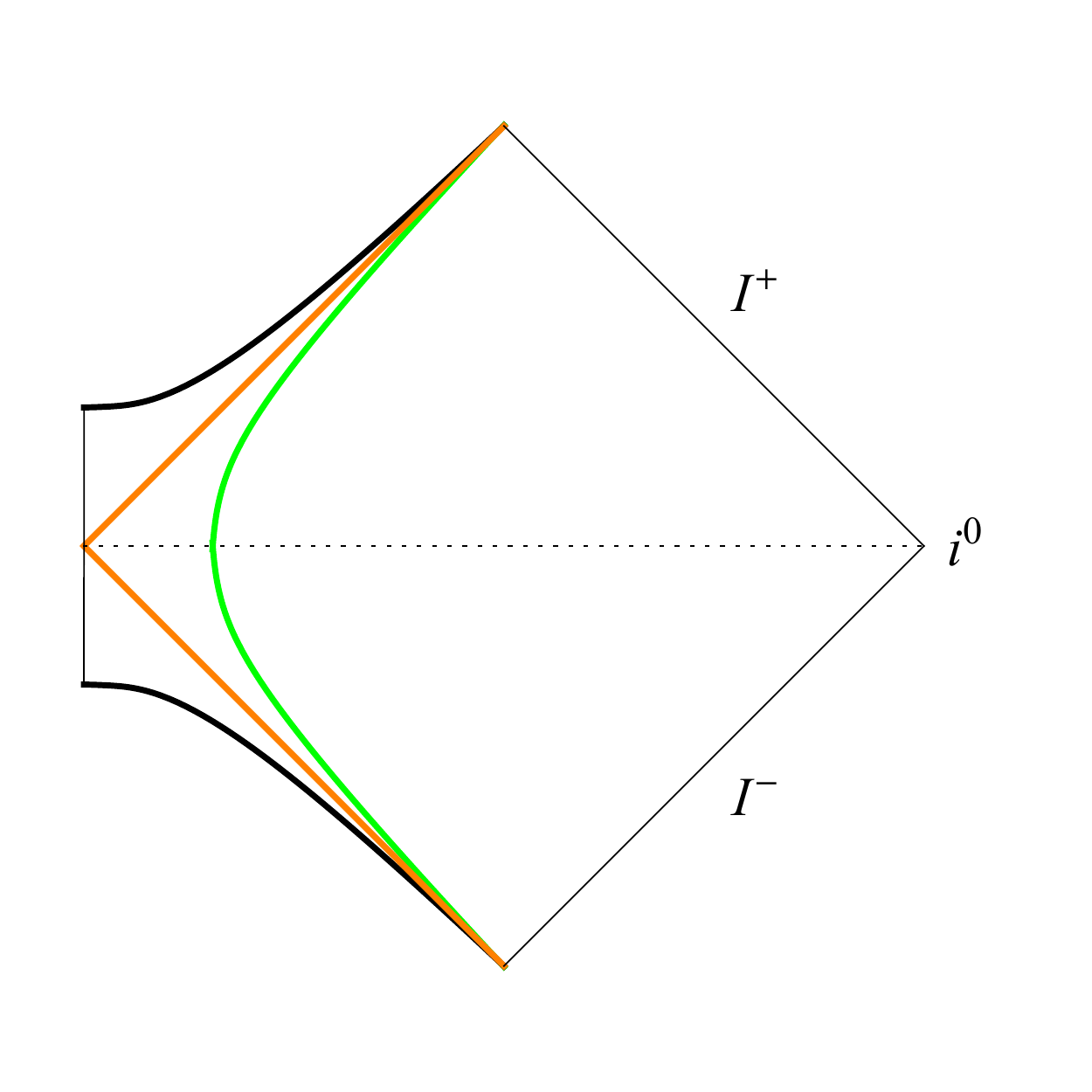} 
\caption{
The geometry described by the metric (\ref{eq:metriconf}) for $h_0=2$, $r_0=4$. 
}
\label{plot1}
\end{figure}

Even if the initial scalar field configuration is smooth, a singularity develops at
sufficiently late times, along the line 
\be
t^2=r^2+(1-h_0/\sqrt{6})r_0^2,
\label{crunchline} \ee 
accompanied by a curvature singularity in spacetime. 
Moreover, a region of trapped surfaces appears around it, whose boundary forms an
apparent horizon, as we shall see in the following. 
The curvature scalar diverges 
for $A(t,r)=0$, leading to eq. (\ref{crunchline}).
This relation determines a surface on which the geometry collapses, as is apparent from
eq. (\ref{eq:metriconf}). 
The singularity has 
a natural interpretation as a generalization of the known `AdS crunch',
associated with dynamical fields in AdS spacetime \cite{Coleman,Higgstory},
to a case in which
the negative vacuum energy is not constant. The spacetime is well defined only in the
$(t,r)$ regions in which $A^2(t,r)>0$. 
In the left plot of fig. \ref{plot1} we depict the location of the singularity as a thick black line
on the $(t,r)$ plane for $h_0=2$, $r_0=4$. 
This line asymptotically converges to the line $t=r$.

The presence of the singularity also leads to the appearance of a region 
of trapped 
surfaces, whose boundary defines an apparent horizon.
Outgoing/ingoing null geodesics, denoted by $r_\pm(t)$, 
define surfaces of areal radii 
$R_\pm(t,r_\pm(t))$. A truly outgoing geodesic results in the growth of the 
area of such a surface, while an ingoing geodesic results in the
reduction of the area. On an apparent horizon, the rate of change of
the area vanishes. The product
\be
\Theta= \frac{dR_+}{dt}  \frac{dR_-}{dt} \label{eq:Thetaspherical} \ee
is a convenient quantity in order to search for a horizon. 
An apparent horizon would appear at the point where $\Theta$ vanishes and subsequently 
changes sign.
In our model, 
outgoing/ingoing geodesics for the metric of eq. (\ref{eq:metriconf}) 
satisfy $dr_\pm /dt=\pm 1$, while the areal radius is $R=Ar$. This gives
\be
\Theta=(A_{,t} r)^2-(A+A_{,r} r)^2,
\label{thetaconformal} \ee
with the subscripts denoting derivatives.
In fig. \ref{plot1} we depict as a blue line the location of the apparent horizon on which 
$\Theta $ vanishes. The highlighted region between the apparent horizon and the singularity is
the region of trapped surfaces. 

The lightcone of an observer at the centre of the instanton, which corresponds to
 $t=r$, is depicted as a thick red line in fig. \ref{plot1}. This surface 
 delineates a region from which light cannot escape. 
  Light rays travel at 45$^o$ towards the right (outgoing) or 
the left (ingoing). When they are emitted from any point above the red line, they 
always end up on the singularity. 
Based on this property, this surface can be characterized as an event horizon. 
Notice that the event horizon appears already at $t=0$, before the emergence of trapped 
surfaces, as expected.

The bubble of true vacuum does not have a sharp surface separating it from the 
bulk space where the field asymptotically approaches the false vacuum at $h=0$. 
One can use as a measure of the size of the bubble the radius at which the field
$h(t,r)$
has a value equal to a fraction $\kappa <1$ of its initial value at the center 
$h(0,0)=\sqrt{6} \arctanh (h_0/\sqrt{6})$.
In fig. \ref{plot1} we depict as a green line the location of this radius for 
$h_0=2 $ and $\kappa=2/3$. It is apparent that the radius grows with a speed
that asymptotically approaches that of light.

The combined picture that emerges is of a spherical bubble of true vacuum that expands
asymptotically at the speed of light. A singularity appears in its interior that 
also expands asymptotically at the speed of light, surrounded by a region of trapped surfaces
whose boundary forms an apparent horizon. 
The singularity and the bubble exterior are separated by an event horizon that 
coincides with the light cone of the observer at the centre of the instanton. 
At finite times, as measured by an asymptotic observer, the singularity is confined behind
the horizon. 
However, the expansion of the bubble surface is not stopped by the collapse of its
interior.
At asymptotically late times the bubble eats up the entire false vacuum, while its interior is swallowed by the singularity.
The picture is consistent with the analysis of \cite{Higgstory,Repeat} for the Higgs potential.

The spacetime can also be depicted through a conformal diagram, shown in the right plot
of fig. \ref{plot1}. In this diagram the time coordinate was allowed to take negative 
values. The bubble surface starts with infinite radius at $t\to -\infty$, moves  
to a finite radius at $t=0$, and subsequently expands again.
The left plot corresponds to the upper half of the right plot. 
It is also apparent that both the `crunch' and the bubble surface 
reach the future null infinity at the same point, swallowing the entire 
false vacuum. The conformal diagram is
reminiscent of that of a Schwarzschild black hole. However, there is a
a crucial difference: the singularity appears not at a single point, but on an entire
spatial surface, similarly to the 
Big Bang singularity. In this sense, the event horizon is a cosmological horizon.

\section{Exact solutions in asymptotically de Sitter space}

In this section we generalize the previous solutions to the case that the false
vacuum has nonzero vacuum energy. Of particular interest is the case of positive energy density,
as the resulting solutions describe tunnelling in de Sitter spacetime, a situation
relevant for inflation. (Similar solutions have been derived for tunnelling from an
AdS false vacuum, with an interepretation in the context of the AdS/CFT correspondence.
See \cite{ads1,ads2} and references therein.)

For a conformally coupled scalar field ($\xi=1/6$) it is possible to find an exact
solution for the potential 
\be
V(h) = \frac{12}{\ell^2} -\frac{1}{4} \lambda h^4
\label{pot12} \ee
with $\lambda>0$.
It describes Euclidean dS space of curvature $48/\ell^2$,
 in coordinates such that the metric takes the form
\be 
g_{\mu\nu}
= \frac{1}{\left( 1+ (r^2+t_{\rm E}^2)/\ell^2 \right)^2} \, \eta_{\mu\nu},
\label{euclideands}\ee
where $ \eta_{\mu\nu}$ is the identity matrix.
The field is given by the relation
\beq 
h(t_{\rm E},r) = \frac{h_0\left(1+(r^2+t_{\rm E}^2)/\ell^2 \right)}{1+(r^2+t_{\rm E}^2)/r_0^2}
\qquad  \hbox{with} \qquad 
r_0= \frac{1}{h_0}\sqrt{\frac{8}{\lambda}}
\label{eq:Fubini2}\eeq
and arbitrary $h_0$.
The action attributed to this configuration can be computed as
\beq S = \int d^4x \sqrt{\det g} 
\left[ \frac12 g^{\mu\nu}(\partial_\mu h)(\partial_\nu h) 
+\frac{1}{2}\xi \Rc\, h^2+ V(h)-V(0) \right].
\label{actionee} \eeq
It is again given by eq. (\ref{actionf}).


A peculiarity of the Euclidean solution (\ref{eq:Fubini2}) is 
that for $r^2+t_{\rm E}^2\to \infty$ the field does not reach zero. 
In this sense the configuration does not start exactly from
the false vacuum. As discussed in detail in \cite{EWeinberg} this is a typical occurrence
in tunnelling from a dS vacuum, which can be considered as a thermal environment at the
characteristic dS temperature. The field can be viewed as being thermally excited 
to a value away from the unstable minimum of the potential, with the tunnelling 
occuring subsequently from that value. 
It must be noted that a consistent picture requires that 
we assume the hierarchy $\ell\gg r_0$, so that the
bubble is nucleated with a surface well within the dS horizon. The asymptotic field value 
is then suppressed by $r_0^2/\ell^2$. Also, realistic potentials include 
small modifications near the origin, which 
generate a locally stable false vacuum at 
vanishing field. These modifications
induce only small corrections to the leading tunnelling solution (\ref{eq:Fubini2}),
while forcing the field to approach a value close to zero for $r^2+t_{\rm E}^2\to \infty$. 
Their main effect is to select a specific value for $h_0$ for the unique instanton solution.

The conformal transformation (\ref{conftrans}) and the field redefinition (\ref{redefinition})
generate a solution for a theory of a scalar field with a canonical kinetic term and a
minimal coupling to gravity ($\xi=0$). 
The potential is given by 
\be
V(h) =\frac{12}{\ell^2}\cosh^4\frac{h}{\sqrt{6}} -9\lambda  \sinh^4\frac{h}{\sqrt{6}}.
\label{pot23} \ee 
The solution of the eom for a Lorentzian signature is
\begin{eqnarray} 
h(t,r)&=&\sqrt{6} \arctanh\frac{h_0}{\sqrt{6}}
\frac{1 + (r^2 - t^2)/\ell^2}{1 + (r^2 - t^2)/r_0^2}
\label{eq:fieldonf2} \\
g_{\mu\nu} =A^2(t,r) \eta_{\mu\nu}
&=& \left[ \frac{1}{\left( 1+ (r^2-t^2)/\ell^2 \right)^2}-\frac{h_0^2/6}{(1+(r^2-t^2)/r_0^2)^2}\right] \eta_{\mu\nu}.
\label{eq:metriconf2}\end{eqnarray}
The continuation to Euclidean signature provides
an $O(4)$-symmetric solution of the eom resulting from the 
action (\ref{actione}) with $\xi=0$ and a
potential given by eq. (\ref{pot23}).
Similarly to above, the field does not reach zero for $r^2+t_E^2 \to \infty$. This is
reflected in the fact that the curvature becomes
\be
\Rc=\frac{48}{\ell^2}\frac{1-\frac{h_0^2}{6}\frac{r_0^6}{\ell^6}}{\left(1-\frac{h_0^2}{6}\frac{r_0^4}{\ell^4}\right)^2}
\label{curvature} \ee
in this limit. Our assumptions that $\ell \gg r_0$ and that additional 
small modifications to the potential
around the origin generate a minimum at this point imply that asymptotically 
the field will approach a value close to zero, while keeping the basic form of
eqs. (\ref{eq:fieldonf2}), (\ref{eq:metriconf2}).
The main effect is that the solution will exist only for a specific value of $h_0$.

\begin{figure}[t!]
\centering
\includegraphics[width=0.5\textwidth]{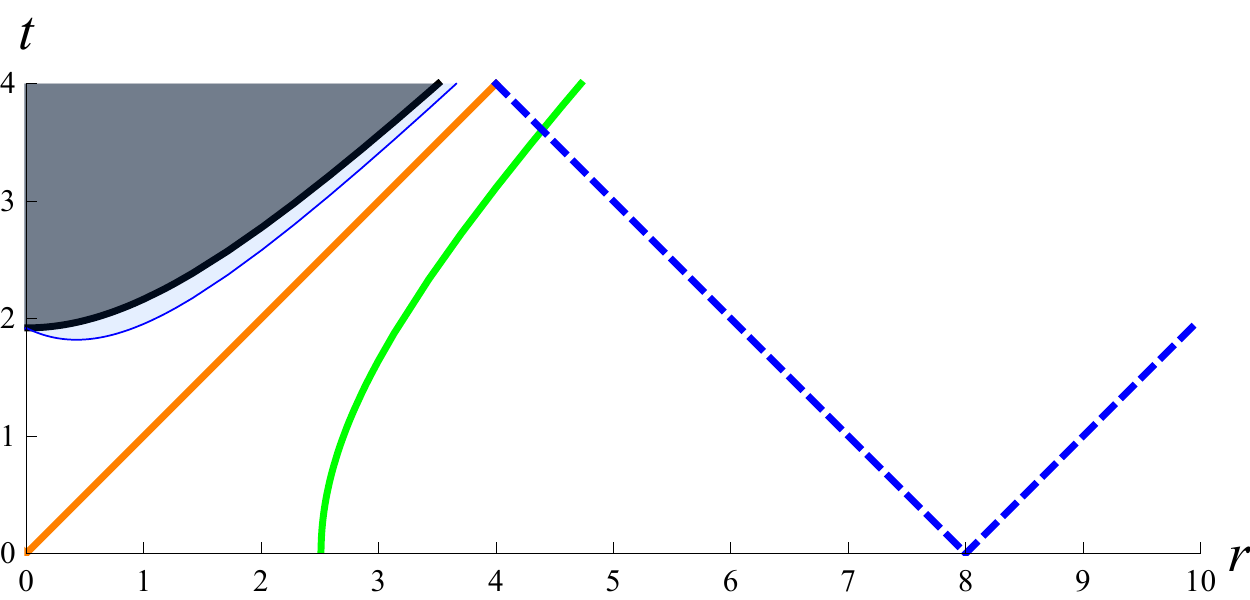} 
\includegraphics[width=0.4\textwidth]{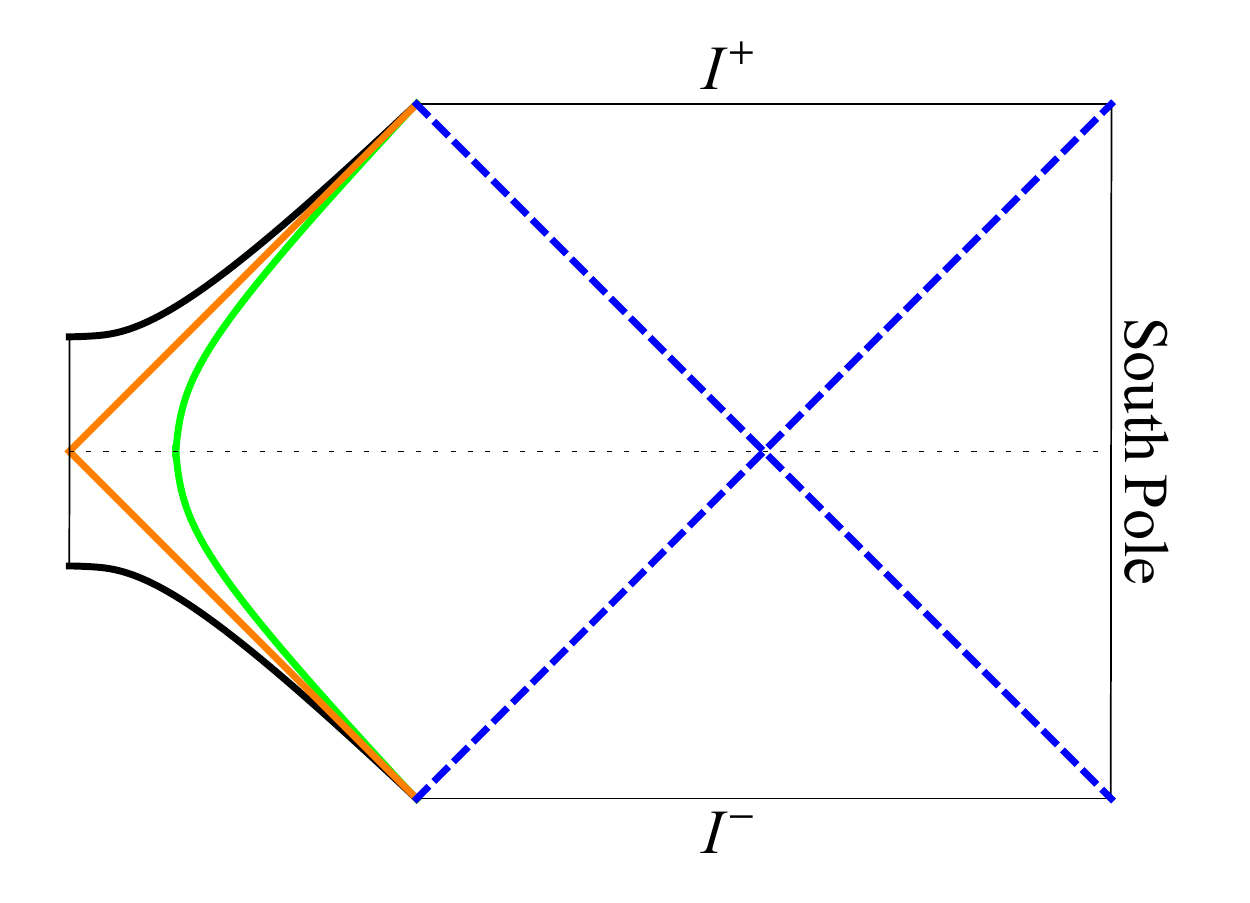} 
\caption{Left plot:
The geometry described by the metric (\ref{eq:metriconf2}) for $h_0=2$, $r_0=4$, $\ell=8$. 
Right plot: The conformal diagram for $\ell \to \infty$.
}
\label{plot2}
\end{figure}

The geometry described by the metric (\ref{eq:metriconf2}) is depicted in fig. \ref{plot2}.
The location of the `crunch' singularity (thick black line), the region of trapped surfaces
(light-blue shaded area), the 
apparent horizon (blue line), the event horizon (red line) and the location of the
bubble surface (green line) are very similar, with small quantitative diffences, to 
those of fig. \ref{plot1}. 
The main difference between the two figures lies in the 
presence of the horizons of the asymptotic dS spacetime. These are depicted as
dashed blue lines. It important to notice that the metric (\ref{eq:metriconf2})
can be written as the dS metric multiplied by a conformal factor. Therefore, it
retains the causal structure of the dS geometry, apart from the region 
affected by the singularity of the conformal factor. As a result, the dS horizons
are expected to persist as horizons of the full geometry, even though the bubble
interior must be viewed as a quasi-AdS spacetime. 

The location of the dS horizons
can be determined by changing from the coordinates 
$(t,r)$ of the metric (\ref{eq:metriconf2}) with $h_0=0$ to Kruskal coordinates. 
In this way, one finds that the horizons correspond to the lines
$t=\pm (\ell -r)$, depicted in fig. \ref{plot1}. 
These lines denote the
limits for the information exchange with the observer on the South pole of the dS spacetime. 
Notice that, in the
left plot of fig. \ref{plot2}, $\ell$ has not
been taken much larger than $r_0$ in order for the details of the plot to be visible.
Around the surface of the
bubble, the geometry switches from quasi-dS to quasi-AdS. 
An extension of the solution to negative values of $t$ allows the construction of the 
conformal diagram, shown in the right plot of fig. \ref{plot2} for $\ell \to \infty$. 

There are three
interesting points that can be deduced from fig. \ref{plot2}. 
\begin{itemize}
\item
The first one 
is the similarity with the picture that emerges from an analysis in the thin-wall limit 
\cite{Higgstory},
based on appropriate junction conditions on the bubble surface \cite{Israel}.
In particular, one notices the crossing of the dS horizon by the
bubble surface at some point during the evolution. 
\item
The second observation is that when the `crunch' and the bubble surface reach the
future null infinity a portion of space remains outside the bubble, so that 
a part of the false vacuum survives. Of course, after the end of inflation the 
relevant conformal diagram is that of fig. \ref{plot1}, which implies that the 
whole of the false vacuum is eventually swallowed by the `crunch'. 
\item
Finally, the similarity of the right plot with the conformal diagram of the 
dS-Schwarzschild geometry is apparent. 
However, there is a crucial difference in 
that the singularity appears not at a point, but on a whole spatial surface,
similarly to the Big Bang singularity. As a result, 
the event horizon depicted by the red line is a cosmological horizon.

\end{itemize}

\section{Numerical solutions for a class of models}

The purpose of this work was to derive analytical solutions of vacuum decay in 
unbounded potentials in order to understand in precise terms the effect on the
bubble surface of the 
`crunch' developing within the nucleated quasi-AdS bubble interior.
Exact solutions were derived for specific potentials, which demonstrated that the 
surface of a sufficiently large bubble keeps expanding, independently of the 
`rolling' of the field down the potential and the appearance of a singularity. 
The results complement the numerical analysis of \cite{Repeat} and reconfirm the 
general picture presented in \cite{Higgstory}.

A variety of other models can be constructed 
in an implicit fashion, resulting in very similar solutions and 
conformal diagrams. It was observed in \cite{Apostolopoulos} that 
a conformally flat metric of the form $g_{\mu\nu}=A^2(w)\,\eta_{\mu\nu}$, with 
$w=r^2-t^2$, and a scalar field configuration obeying 
\be
h'^2(w)=\frac{2}{A^2(w)}\left(2A'^2(w)-A(w)A''(w) \right)
\label{scalarw} \ee
satisfy the Einstein equations for a minimally coupled scalar theory
with a potential given implicitly as 
\be
V(w)=\frac{4}{A^4(w)}\left( w A'^2(w)+3 A(w)A'(w)+3w A''(w)\right).
\label{potw} \ee
This allows the construction of models directly in the Einstein frame by 
selecting the desired $A(w)$,
solving eq. (\ref{scalarw}) for $h(w)$, inverting this function and deriving 
the function $V(h)$ through substitution in eq. (\ref{potw}). Even though 
finding analytical
solutions is not easy, a numerical implementation of the above procedure is 
straightforward. An interesting class of solutions 
has metrics of the form
\be
g_{\mu\nu} =A^2(t,r) \eta_{\mu\nu}
= 
\left[ \frac{1}{\left( 1+ (r^2-t^2)/\ell^2 \right)^2}\right]
\left[ 1-\frac{h_0^2/6}{(1+(r^2-t^2)/r_0^2)^s}\right] \eta_{\mu\nu},
\label{finalmetric} \ee
with $s$ an integer larger than 2. The metric is similar to (\ref{eq:metriconf2}). 
It includes one conformal factor that 
corresponds to an exact dS spacetime and another that induces a `crunch' singularity.
For $s >2$ the dS metric is approached in the limit $r\to \infty$ faster than 
in the case of the metric (\ref{eq:metriconf2}). 

\begin{figure}[t!]
\centering
\includegraphics[width=0.4\textwidth]{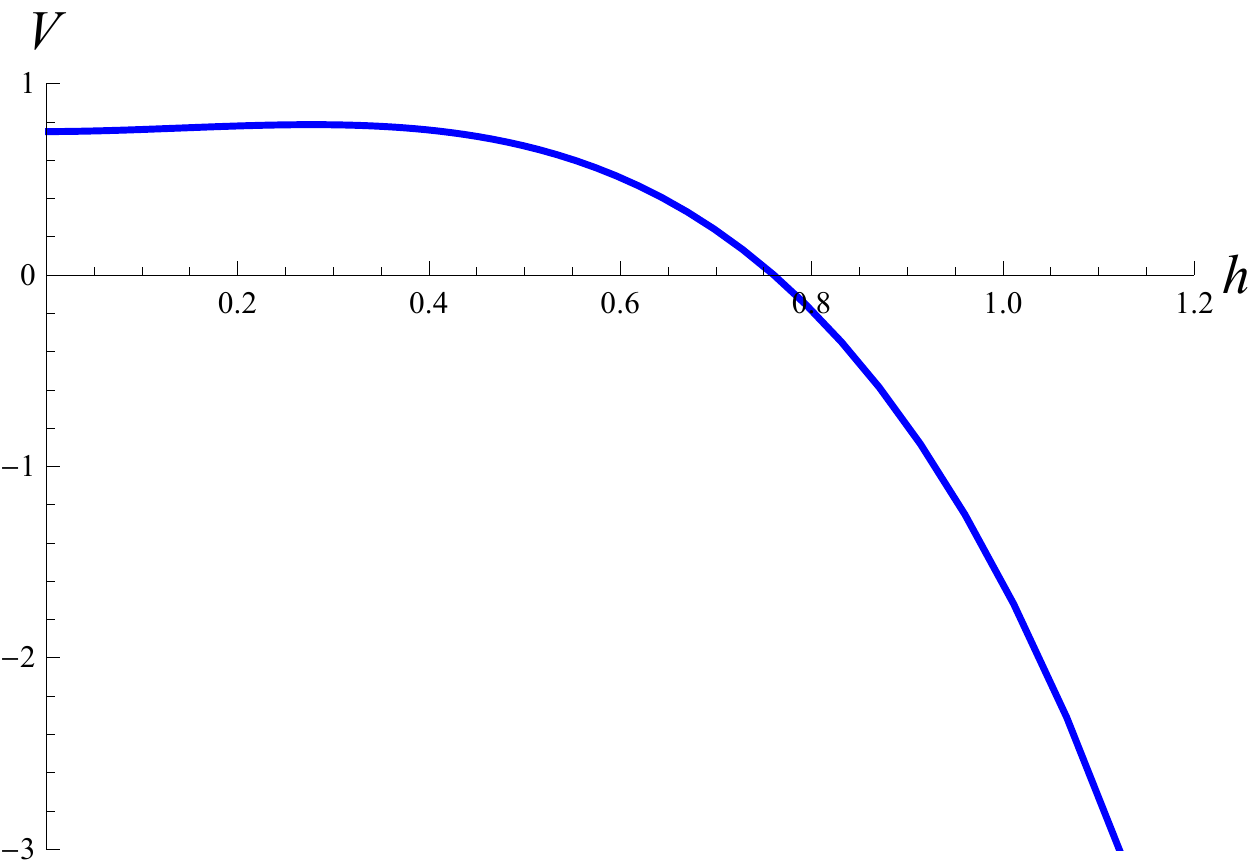} 
\hspace{2cm}
\includegraphics[width=0.4\textwidth]{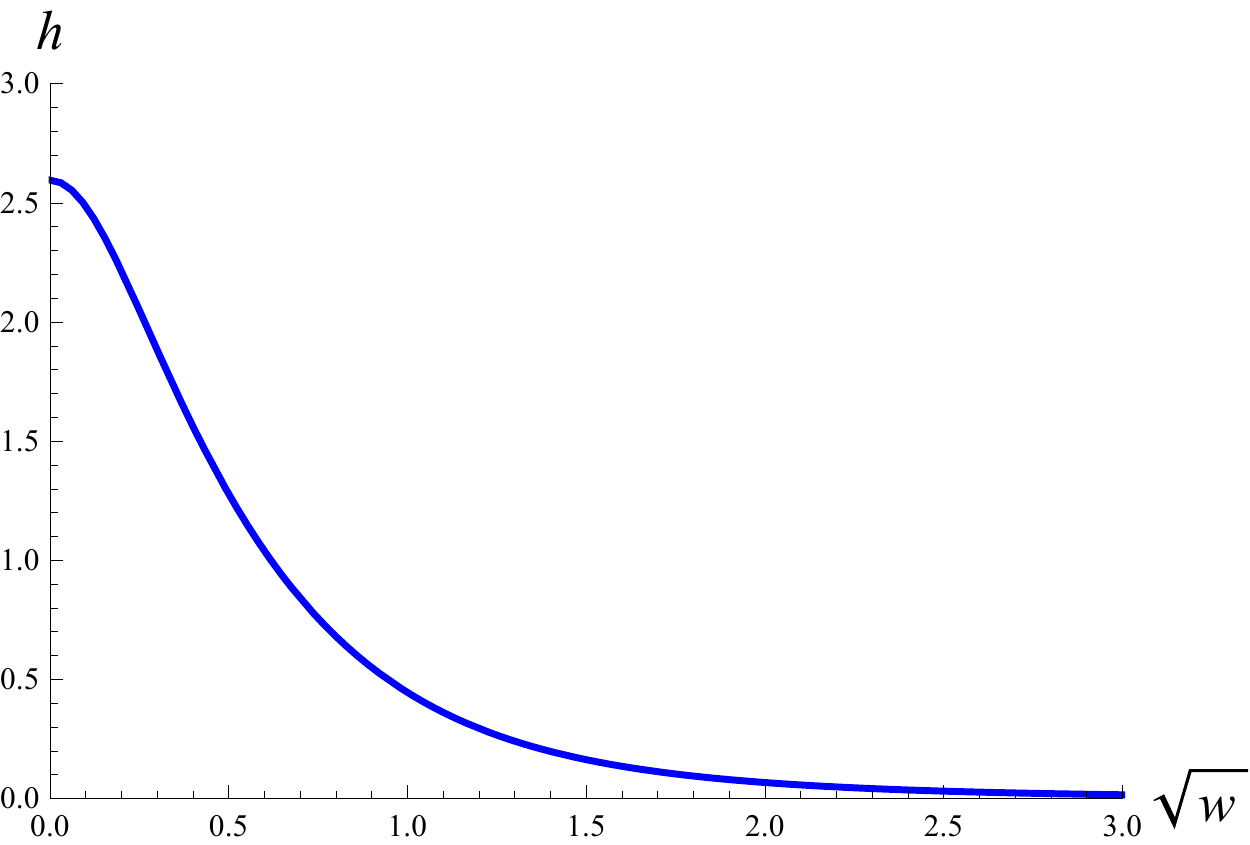} 
\caption{
The potential $V(h)$ and the solution for the field $h(w)$, for the metric
of eq. (\ref{finalmetric}) with $r_0=1$, $\ell=4$, $h_0=2$ and $s=4$.
}
\label{plot3}
\end{figure}

The solution 
of eq. (\ref{scalarw}) describes a field that vanishes for $w\to \infty$. The resulting
configuration 
for an Euclidean signature describes an instanton that drives the tunnelling from the 
false vacuum at the origin to the unstable region of the potential.
This configuration is depicted in the right plot of fig. \ref{plot3} as a function of 
$\sqrt{w}=\sqrt{r^2+t_E^2}$. The potential is given implicitly through eq. (\ref{potw}) and is 
depicted in the left plot of fig. \ref{plot3}. It has a shallow minimum at the origin,
separated by a barrier from the large-field region where it becomes unbounded from below.
This numerical solution 
confirms the validity of the discussion in the previous
section and refs. \cite{Higgstory,Repeat}. The bubble expands continuously, despite the 
appearance of the `crunch' in its interior, crosses the horizon of the external dS geometry,
and eventually reaches asymptotic null infinity. 

\section*{Acknowledgments}
I would like to thank P. Apostolopoulos, C. Charmousis, 
I. Papadimitriou, A. Strumia and R. Troncoso for useful discussions.
The research of N. Tetradis was supported by the Hellenic Foundation for
Research and Innovation (H.F.R.I.) under the “First Call for H.F.R.I.
Research Projects to support Faculty members and Researchers and
the procurement of high-cost research equipment grant” (Project
Number: 824).


\end{document}